\documentclass[prl,twocolumn,floatfix]{revtex4}

\usepackage{graphicx}
\usepackage{color}
\usepackage[normalem]{ulem}
\usepackage{braket}
\usepackage{amsmath}
\usepackage{verbatim}
\usepackage{amssymb}

\newcommand{\be}{\begin{equation}}
\newcommand{\ee}{\end{equation}}
\newcommand{\ba}{\begin{eqnarray}}
\newcommand{\ea}{\end{eqnarray}}

\begin{document}

\title{Searching for quantum speedup in quasistatic quantum annealers}

\author{Mohammad H.~Amin }
\affiliation{D-Wave Systems Inc., 3033 Beta Avenue, Burnaby BC
Canada V5G 4M9}
\affiliation{Department of Physics, Simon Fraser
University, Burnaby, BC Canada V5A 1S6}

\begin{abstract}
We argue that a quantum annealer at very long annealing times is likely to experience a quasistatic evolution, returning a final population that is close to a Boltzmann distribution of the Hamiltonian at a single (freeze-out) point during the annealing. Such a system is expected to correlate with classical algorithms that return the same equilibrium distribution. These correlations do not mean that the evolution of the system is classical or can be simulated by these algorithms. The computation time extracted from such a distribution reflects the equilibrium behavior with no information about the underlying quantum dynamics. This makes the search for quantum speedup problematic.
\end{abstract}

\maketitle

Quantum annealing (QA) \cite{Finnila,Nishimori,Santoro,Brooke99,Farhi01} is a means for solving optimization problems using quantum mechanics. Recently, QA processors with up to more that a thousand qubits have been developed \cite{Johnson11} and tested by many independent research groups \cite{Boixo13,Ronnow14,Hen15a,Hen15b,Vinci15,BoixoNP14,Albash15,Boixo15,Perdomo14,Venturelli14}. The processors are designed to implement a transverse Ising Hamiltonian:
 \ba
 H(s) &=& - A(s)\sum_{i}\sigma^x_i + B(s){\cal H}_P, \\
 {\cal H}_P &=& \sum_{i}h_i\sigma^z_i + \sum_{i<j}
 J_{ij}\sigma^z_i\sigma^z_j, \label{HP}
 \ea
where $\sigma^{x,z}_i$ are Pauli matrices acting on qubit $i$, $s = t/t_a$, $t$ is time, $t_a$ is the annealing time, $h_i$ and $J_{ij}$ are tuneable dimensionless parameters, and $A(s)$ and $B(s)$ are monotonic functions such that $A(0) \gg B(0) \approx 0$ and $B(1) \gg A(1) \approx 0$ (see Fig.~\ref{fig2}a). A successful computation yields the ground state or an acceptable low energy eigenstate of ${\cal H}_P$.

By now, the presence of quantum mechanics \cite{Johnson11,Boixo13}, including entanglement \cite{Lanting14} and computational benefit of multiqubit tunneling \cite{Boixo15} has been established in the QA processors. However, whether or not quantum mechanics can lead to any scaling advantage ({\rm quantum speedup}) over available classical algorithms remains an open question. Recently, there have been many attempts to detect signatures of quantum speedup in D-Wave Two (DW2) QA processors \cite{Ronnow14,Hen15a,Hen15b,King15}. In these studies the processor's performance is compared with some classical algorithms, such as simulated annealing (SA), quantum Monte Carlo simulation of QA (SQA), or other algorithms. The computation time $t_c$ is typically expressed as a function of the success probability, which is usually taken to be the final ground state probability $P_0$. Here, we define
 \be
 t_c = t_a/P_0 \label{t_c2},
 \ee
which is asymptotically equivalent to the definition used in previous publications \cite{Ronnow14,Hen15a,Hen15b,King15}. For a classical solver such as SA or SQA, $t_a$ is taken to be proportional to the number of sweeps (number of iterations in which all spins are updated in a single anneal) \footnote{This is $1/N$-times smaller than the total computation time if only a single-core processor is used. This factor is incorporated to include the possibility of classical parallelism (for more detailed discussion see Ref.~\cite{Boixo15}).}.

\begin{figure}[t]
    \includegraphics[width=6.5cm]{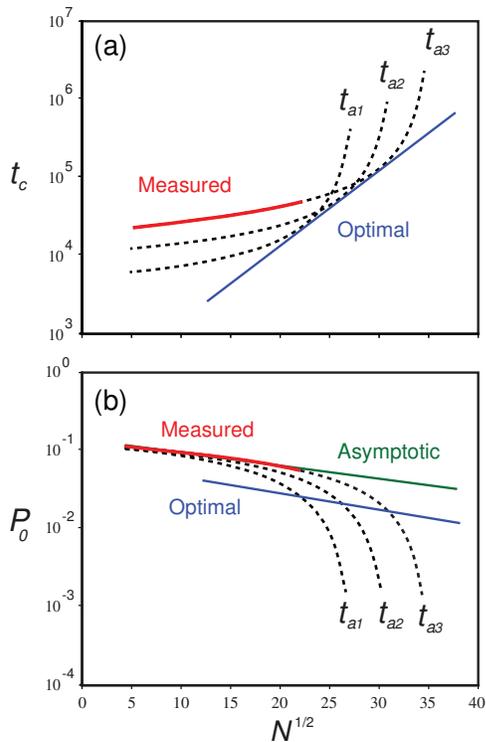}
\caption{(a) A sketch of computation time as a function of $\sqrt{N}$ for different annealing times $t_{a1}<t_{a2}<t_{a3}$. Similar behavior has been observed in SA and SQA and conjectured for QA \cite{Ronnow14,Hen15a}. The blue line shows scaling of an optimized solver. The red line represents the region accessible by a quantum annealer with $t_a \geq t_{a3}$ and $N \leq 512$. (b) The success probability for the same system. The optimal curve in (b) is plotted by connecting the corresponding tangency points in (a). We have assumed $T>0$ to correspond better with a realistic quantum annealer. At $T=0$, the asymptotic probability is flat at $P_0 = 1$. }
    \label{fig1}
\end{figure}

\begin{figure}[t]
    \includegraphics[width=6cm]{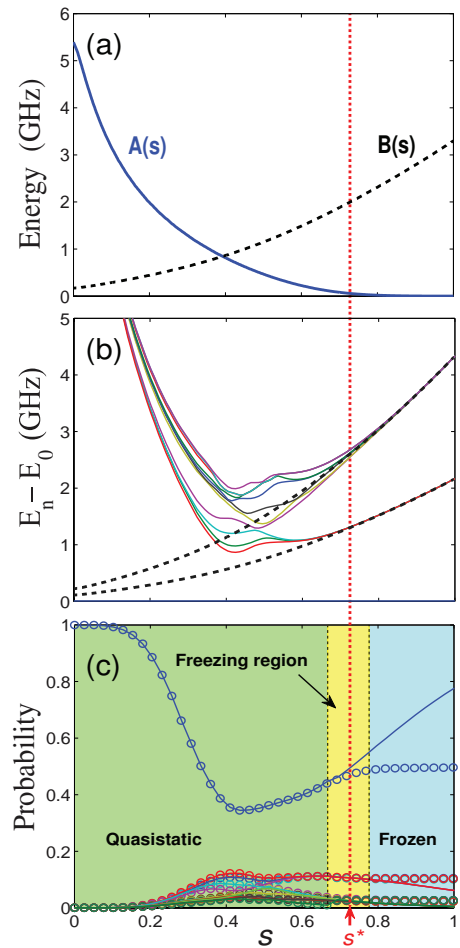}
\caption{(a) Envelope functions $A(s)$ and $B(s)$ for a DW2 quantum annealer. (b) The lowest 12 energy levels of a randomly generated 16 qubit problem. Dashed black lines are classical energies. Notice that the quantum energy eigenvalues are close to the classical ones at $s^*$. (c) Occupation probabilities during the annealing calculated using Redfield formalism (circles) and Boltzmann distribution (solid lines), assuming $T=40$ mK and $t_a = 20$ $\mu$s. All probabilities follow the Boltzmann distribution in the quasistatic region (green) until they start freezing in the freezing region (yellow) and stay constant in the frozen region (blue). All final probabilities are close to the Boltzmann probabilities at the freeze-out point $s^*$, marked by the vertical (red) dashed line. }
    \label{fig2}
\end{figure}

One hopes to determine how $t_c$ scales with the problem size $N$ and whether the quantum annealer provides a better scaling compared to classical solvers. For problems with essentially 2 dimensional structure, such as the Chimera graph (the native graph of DW2), $t_c$ is expected to be an exponential function of $\sqrt{N}$ (i.e., the tree-width of the graph). The slope of $\log(t_c)$ versus $\sqrt{N}$, therefore, provides the coefficient in the exponent. Quantum speedup then means that the quantum annealer provides a smaller slope than other classical solvers.
In practice, complications arise due to the dependence of the scaling curves on $t_a$. Figure \ref{fig1}a shows schematically a behavior commonly observed from some annealing algorithms, such as SA or SQA. The slope of the curves are small for small $N$, because $t_a$ is typically too long for such easy problems. The slope increases abruptly at large $N$ reflecting the fact that the chosen $t_a$ is not long enough to find a solution. This makes the asymptotic scaling of the algorithm unclear. To deal with this problem, Ref.~\cite{Ronnow14} suggests optimizing all of the annealers at each size, i.e., finding the optimal $t_a$ that minimizes $t_c$ for each $N$. The resulting ``optimal'' curve will be independent of a particular choice of $t_a$ (the blue line in Fig.~\ref{fig1}).

In practice, QA processors come with their own limitations in terms of control parameters or available annealing schedules. For example, DW2 annealing curves, $A(s)$ and $B(s)$, are fixed and the minimum available $t_a$ is 20 $\mu$s, which is typically too long rendering suboptimal computation. In the absence of an optimal scaling, it was suggested \cite{Ronnow14,Hen15a} to determine an upper bound for quantum speedup using a suboptimal scaling. Suppose Fig.~\ref{fig1}a is the $t_c$ obtained from a quantum annealer with the minimum allowed annealing time $t_{a3}$ and maximum 512 qubits ($\sqrt{N} < 23$). The red curve in Fig.~\ref{fig1}a represents the experimentally accessible region. It is clear that the slope of the red curve in Fig.~\ref{fig1}a is not a representative of the asymptotic performance. If we were allowed to decrease $t_a$ so that optimization becomes feasible, it would have been possible to obtain the blue (optimal) curve, which has more slope than the red one. Therefore, the slope of the measured (red) curve provides a lower bound for that of the optimal (blue) curve.

To acquire more intuition, we plot in  Fig.~\ref{fig1}b the success probability $P_0$, by inverting each curve in Fig.~\ref{fig1}a and shifting vertically (since $\log P_0 = \log t_a - \log t_c$). The curves in Fig.~\ref{fig1}b tend to overlap at small $N$, demonstrating weak dependence of $P_0$ on $t_a$ for small $N$ (or long $t_a$). Such a weak dependence, commonly observed in both QA and classical algorithms, is an indication of {\em quasistatic} behavior. In thermodynamics, a system is called quasistatic when its time-dependence is very slow compared to its relaxation time so that it stays near the equilibrium state at all times. Typically, in QA the system will follow equilibrium distribution up to some point but then starts deviating from equilibrium and the probabilities will all freeze shortly after. If the time between the deviation and freezing is short, the final probabilities will be close to the equilibrium probability distribution at a single (freeze-out) point within the freezing region. To demonstrate this behavior numerically, we consider a 16 qubit problem with $h_i$ selected from $\pm 1/3$ and $J_{ij}$ from $\pm 1/3$ or $-1$ uniform randomly. Figure \ref{fig2}b shows (solid lines) the 12 lowest energy eigenvalues of $H(s)$ obtained with realistic $A(s)$ and $B(s)$ plotted in \ref{fig2}a. The dashed black lines represent the corresponding classical energies, i.e., eigenvalues of $B(s) {\cal H}_P$. We calculate the occupation probabilities using the open quantum (Redfield) master equation discussed in Refs.~\cite{Amin_BR_09,Johnson11,Boixo15}. Similar master equations have proven to provide good qualitative and quantitative descriptions of superconducting QA processors \cite{Harris10,Johnson11,Boixo13,Boixo15,Vinci15}. We assume an ohmic environment at equilibrium temperature $T=40$ mK with dimensionless coupling constant $\eta = 0.24$. The same model with the same parameters was successfully used to explain the experimental data in Ref.~\cite{Boixo15}. Here, we have chosen a larger than normal temperature to deliberately populate the excited states, otherwise the desired effects would not have been visible. Circles in Fig.~\ref{fig2}c represent the occupation probabilities of the lowest 12 eigenstates during the evolution. We have also plotted the equilibrium probabilities (solid lines) using Boltzmann distribution: $P_n^{\rm B}(s) {=} Z^{-1} e^{-E_n(s)/k_BT}$, where $E_n(s)$ is an instantaneous eigenvalue of $H(s)$, $k_B$ is the Boltzmann constant, and $Z {=} \sum_n e^{-E_n(s)/k_BT}$ is the partition function. As can be seen in the figure, the probabilities closely follow $P_n^{\rm B}(s)$ up to almost 2/3 of the evolution (green region). As $s{\to}1$, $A(s)$ becomes smaller making the thermal relaxation slower. When the relaxation becomes too slow, the system stops following equilibrium and the probabilities start deviating from Boltzmann until they all saturate (freeze). The saturations happen within a freezing (yellow) region. If the freezing region is narrow enough, then the final probabilities will be close to the Boltzmann probability distribution at a single freeze-out point $s=s^*$, marked by the red vertical dotted line in Fig.~\ref{fig2}.

Typically, $s^*$ depends weakly on $t_a$. Let $\gamma(s)$ denote the dominant relaxation rate at $s$. As $s{\to}1$, $\gamma(s)$ vanishes exponentially due to exponential decay of $A(s)$, i.e., $\gamma(s) {\sim} \gamma_0 e^{-\alpha s}$ where $\gamma_0$ and $\alpha$ are problem dependent constants. The freeze-out happens when $\gamma(s)$ becomes too small compared to $t_a^{-1}$, therefore no thermal transition can happen within the annealing time. This mean $\gamma_0 e^{-\alpha s^*} \approx t_a^{-1}$, which yields $s^* \approx \ln(\gamma_0 t_a)/\alpha$ or $P_0(t_a) \approx P_0^{\rm B}[\ln(\gamma_0 t_a)/\alpha]$. Expanding to linear order around $s^*$ gives
 \be
 P_0(t_a) \sim (\kappa/\alpha)\ln(\gamma_0 t_a), \label{P0ta}
 \ee
where $\kappa$ is the expansion coefficient. This weak logarithmic dependence is responsible for the asymptotic behavior illustrated in Fig.~\ref{fig1}b.

Notice in Fig.~\ref{fig2}b that the quantum energy eigenvalues (solid lines) at $s=s^*$ are not far away from the classical ones (dashed lines), although earlier they are very different. The final probabilities are therefore close to a classical Boltzmann distribution, even though the dynamics during the evolution are not classical. This can explain correlations with SA as reported in, e.g., Ref.~\cite{Hen15a}.

\begin{figure}[t]
    \includegraphics[width=7cm]{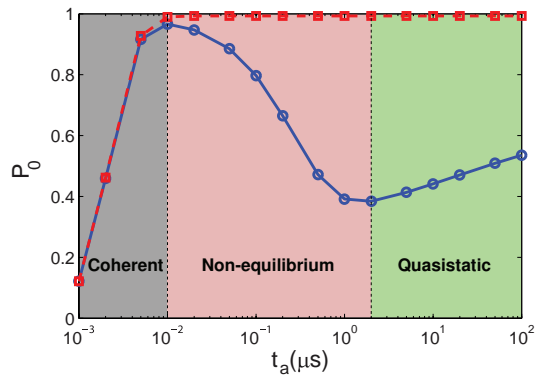}
\caption{Ground state probability as a function of $t_a$ for the problem instance studied in Fig.~\ref{fig2}. The circles are calculated using the Redfield master equation calculation with the same parameters as in Fig.~\ref{fig2}c. The squares are results of closed system calculations. Three regions are distinguished: {\em Coherent}, where the environment does not have enough time to affect the system hence the open and closed system probabilities coincide. {\em Non-equilibrium}, in which the environment starts to occupy the excited states during the evolution, but does not have enough time to establish equilibrium. {\em Quasistatic}, where the evolution is so slow that the system follows equilibrium during most of the evolution (as in Fig.~\ref{fig2}c) and $P_0$ depends on $t_a$ according to (\ref{P0ta}). }
    \label{fig3}
\end{figure}

It has been conjectured that QA can become advantageous over thermal annealing or SA for problems that have thin but tall barriers in their energy landscapes \cite{Brooke99}. The reason is tunneling through such barriers becomes more efficient than thermal excitation over the barrier. This is certainly a dynamical effect and not an equilibrium property. If the source of a possible quantum speedup is dynamical, as conjectured, then searching for the speedup requires access to such dynamics. In general, an equilibrium population is only a function of the energy eigenvalues and therefore is independent of the dynamics. For a quasistatic quantum annealer with a sharp freeze-out at $s=s^*$, the final ground state population is close to $P_0^{\rm B}(s^*)$, which is only a function of $E_n(s^*)$. It is clear that from such probabilities it is not possible to determine how fast the distribution is established. For example, consider a hypothetical quantum annealer that can return a Boltzmann distribution of $H(s^*)$ within a {\em constant} time $t_a = O(1)$ (independent of $N$) that is much faster than any other classical solvers and obviously scales better than all of them. Since sampling from a Boltzmann distribution is a very hard computational problem, this hypothetical quantum annealer (although may never exist) can provide an incredible quantum speedup for such a hard problem. Nevertheless, when assessed based on a suboptimal $t_c$, it may scale worse than some optimized classical solvers, merely due to the scaling of $P_0^{\rm B}(N)$.

We now can return to our original question of whether it is possible to determine a bound for quantum speedup from suboptimal data. If the measured slope in Fig.~\ref{fig1}a provides a lower bound for the optimal slope, as it appears in that figure, then one may know with certainty when quantum speedup is not possible. A measured (suboptimal) slope being larger than the classical slopes means that the optimal slope would be even larger. Therefore the optimized QA would scale even worse and it can never be able to provide quantum speedup. On the other hand, if the measured slope is smaller than classical, then no conclusion can be made because there is always a possibility that the optimal slope be larger than classical. In other words, it is possible to ``rule out'' quantum speedup from a suboptimal performance, but it is not possible to prove it. Based on what we discussed before, if the suboptimal probabilities do not provide information about the quantum dynamics, then it should not be possible to even put a bound on quantum speedup, which seems to contradict the above argument. To resolve this, we notice that the situation depicted in Fig.~\ref{fig1}a is based on assumptions which may not hold in general for QA. For example, it is assumed that the slope of each curve at a fixed $t_a$ is a monotonically increasing function of $N$, or that $t_c$ increases with $t_a$ at small $N$, but decreases at large $N$. It is also assumed (implicitly in Fig.~\ref{fig1}a, but explicitly in Fig.~\ref{fig1}b) that the probability $P_0$ is a monotonically increasing function of $t_a$. Most of these assumptions are based on observations from classical solvers such as SQA (see e.g., supplementary information of Ref.~\cite{Ronnow14}) and may even apply to most classical solvers. Establishing that they also hold in general for QA requires a proof. It is, however, possible to disprove them by a counterexample, as we do next.

\begin{figure}[t]
    \includegraphics[width=6cm]{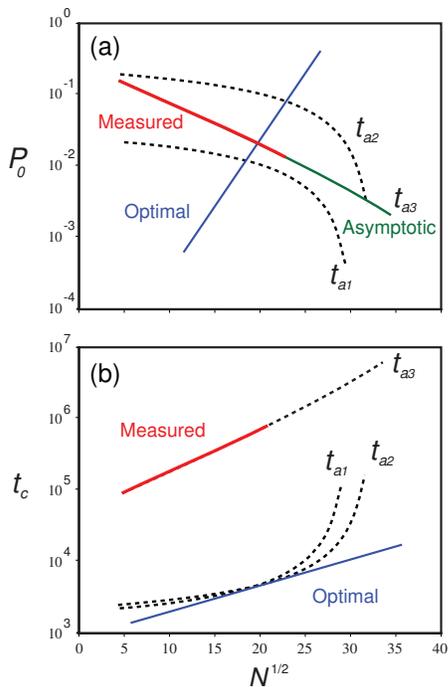}
\caption{(a) $P_0$ and (b) $t_c$ as a function of $\sqrt{N}$ assuming a non-monotonic $P_0(t_a)$. The probabilities at $t_{a2}$ exceed the asymptotic line. The optimal curve in (a) is plotted by connecting the corresponding tangency points in (b).}
    \label{fig4}
\end{figure}

As a counterexample, we examine the last assumption mentioned above, i.e., $P_0(t_a)$ is a monotonically increasing function. We calculate $P_0$ as a function of $t_a$ for the 16 qubit instance of Fig.~\ref{fig2} using the Redfield model that we used before with the same realistic parameters that have shown agreement with experiment \cite{Boixo15}. Circles (squares) in Fig.~\ref{fig3} are results of the master equation calculations with (without) coupling to the environment. It is clear that the open system probability behaves non-monotonically. For short $t_a$, the environment does not have enough time to excite the system from the ground state, therefore, the open and closed system probabilities coincide (gray region). For intermediate $t_a$, the environment starts to populate the excited states through a non-equilibrium dynamics and the probabilities decrease (pink region). At long $t_a$, the system starts observing quasistatic evolution with an equilibrated final population and a $P_0$ that increases according to Eq.~(\ref{P0ta}). The nonmonotonic $t_a$-dependence is expected to be more generic than the chosen example. Since in QA the system starts in the ground state, the thermal excitation populates the excited states from bottom up (unlike the top down relaxation in SA). By decreasing $t_a$ beyond the relaxation time, the thermal processes will not have enough time to excite the system, thus $P_0$ will increase. When annealing becomes so fast that the non-adiabatic excitations become possible, $P_0$ will start decreasing again. Similar behavior has recently been observed in SQA \cite{Steiger}.

A non-monotonic $P_0(t_a)$ may result in a performance very different from what is predicted in Fig.~\ref{fig1}a. If we allow probabilities to exceed the asymptotic probability for small $t_a$ in Fig.~\ref{fig1}b, then we may obtain Fig.~\ref{fig4}a, which leads to a $t_c$ plotted in Fig.~\ref{fig4}b. The plotted optimal line in Fig.~\ref{fig4}b has now a smaller slope than the asymptotic one, contradicting the argument used to bound quantum speedup. It should be emphasized that the behavior depicted in Fig.~\ref{fig4} serves only as a counterexample and we do not claim it to be generic for QA. Other situations may arise that could be different from both Figs.~\ref{fig1} and \ref{fig4}.

It is important to mention that the final probabilities of a quantum annealer is not always an equilibrium population, even for long $t_a$. Since the relaxation rates between different energy levels change differently with time, they may freeze at different points during the annealing, leading to a distributed freeze-out (a wide freezing region in Fig.~\ref{fig2}c). In such cases, it is not possible to identify a single freeze-out point at which the equilibrium population gives the final population. Such situations are likely to happen when the problem has a complicated energy landscape involving numerous valleys with large barriers between them. These problems are expected to have final populations that are more sensitive to $t_a$ and do not correlate with equilibrated QMC simulations. In Ref.~\cite{Katzgraber}, it was shown that random problems on the Chimera graph have no spin glass phase transition at any nonzero temperature. The lack of spin glass phase transition is an indication that the classical energy landscape of the problem is not very complex, thus equilibration may be easy. This signifies the importance of the problem selection for any exploration of the role of quantum dynamics in QA, as correctly pointed out in Ref.~\cite{Katzgraber}.

To summarize, we have shown that a quasistatic evolution with long annealing time can mask the underlying quantum dynamics in a quantum annealer. The final population of such an annealer is likely to be close to a Boltzmann distribution at a single freeze-out point. It is therefore expected to correlate well with a quantum Monte Carlo simulation equilibrated at the same point with a correct temperature. Reference \cite{BoixoNP14} provides indications of such correlations, although QMC was used as an annealing algorithm (SQA) with time-dependent Hamiltonian (not at a fixed $s^*$), and the used $T$ was smaller than the real temperature. This may explain why correlations did not persist when considering excited states \cite{Albash15}. Correlation with SA is also possible  (see e.g., Ref.~\cite{Hen15a}), but requires the quantum eigenenergies to be close to the classical ones at the freeze-out point. Other semiclassical models may also correlate with a quasistatic quantum annealer if they reach the same equilibrated population (see e.g., Ref.~\cite{SSSV}). These correlations are signatures of a quasistatic behavior and do not mean that the dynamics of the quantum annealer can be simulated by such algorithms. The lack of access to the relevant quantum dynamics makes the search for quantum speedup in a suboptimal quantum annealer Problematic. Since one cannot neither prove nor rule out the possibility of quantum speedup, no conclusion can be made unless the optimization procedure proposed in Ref.~\cite{Ronnow14} is done properly. We should emphasize that our argument does not undermine the importance of benchmarking in assessing the performance of a quantum annealer, even if suboptimal, nor do we claim that a quasistatic quantum annealer is not useful. Indeed, providing samples from a Boltzmann distribution is a very hard computational problem with many applications especially in machine learning. However, different benchmarking strategies are required to assess QA for such applications \cite{Katzgraber15}.

We thank T.\ Albash, E.\ Andriyash, S.\ Boixo, I.\ Hen, S.\ Isakov, W.\ Kaminsky, A.\ King, D.\ Lidar, C.\ McGeoch, H.\ Neven, J.\ Raymond, J.\ Rolfe, A.\ Roy, A.\ Smirnov, and R.\ de Sousa, for useful discussions and comments on the manuscript.

\end{document}